\documentclass{iopart}

\pdfoutput=1

\usepackage{epsfig}
\usepackage{iopams}
\usepackage{hyperref}
\usepackage{amssymb}
\usepackage{bm,bbm}
\usepackage{color}
\def\sgn{\mathop{\textrm{sgn}}}

\newcommand{\beq}{\begin{equation}}
\newcommand{\eeq}{\end{equation}}
\newcommand{\beqarray}{\begin{eqnarray}}
\newcommand{\eeqarray}{\end{eqnarray}}

\newcommand{\eq}[1]{Eq.~\eqref{#1}}

\newcommand{\eqref}[1]{(\ref{#1})}

\begin{document}

\title{Helical spin texture of surface states in topological superconductors}

\date{\today}

\author{P. M. R. Brydon}
\ead{pbrydon@umd.edu}
\address{Condensed Matter Theory Center, Department of Physics,  The University
  of Maryland, College Park, MD 20742-4111, USA}

\author{Andreas P. Schnyder}
\ead{a.schnyder@fkf.mpg.de}
\address{Max-Planck-Institut f\"ur Festk\"orperforschung,
  Hei\ss{}enbergstrasse 1, D-70569 Stuttgart, Germany}

\author{Carsten Timm}
\ead{carsten.timm@tu-dresden.de}
\address{Institute of Theoretical Physics, Technische Universit\"at
  Dresden, D-01062 Dresden, Germany}

\begin{abstract}
Surface states of topological noncentrosymmetric
superconductors exhibit intricate helical spin textures, i.e., the
spin orientation of the surface quasiparticles is coupled to
their momentum. Using quasiclassical theory, we study the spin
polarization of the surface states as a function of the spin-orbit
interaction and superconducting pairing symmetry.
We focus on two- and three-dimensional fully gapped and nodal
noncentrosymmetric superconductors. For the case of
nodal systems, we show that the spin polarization of the topological
flat bands is controlled by the spin polarization of the bulk normal
states at the bounding gap nodes. We demonstrate that the zero-bias
conductance in a magnetic tunnel junction can be used as an
experimental test of the surface-state spin polarization.
\end{abstract}

\date{\today}


\maketitle

\tableofcontents

\section{Introduction}

Topological superconductors are characterized by  protected zero-energy surface
states that arise because of the nontrivial topology of the bulk wavefunctions
in momentum
space~\cite{schnyderPRB08,kitaev22,schnyderAIP,hasanKaneRMP,ryuNJP2010,
qiZhangRMP11}.
These surface states are associated with one out of several topological
invariants. There is currently an intense research effort aimed
at identifying topological superconductors, but unambiguous examples
of such phases have not yet been established.
Noncentrosymmetric superconductors (NCSs), characterized by
mixed-parity pairing and strong spin-orbit coupling, have been extensively
investigated as possible candidate materials for topological
superconductivity~\cite{satoPRB06,yadaPRB2011,schnyder2011,schnyder2012,brydon2011,tanakaPRL2010,dahlhaus2012,matsuuraNJP2013,satoFujimoto2009,qiZhang2010,yipLowTemp10,tanakaNagaosaPRB09,satoPRB2011,tanakaJPSJ2012,bauerSigristbook},
the most prominent examples being
Li$_2$Pd$_x$Pt$_{3-x}$B~\cite{yuanPRL06,nishiyama07}, BiPd~\cite{joshiPRB11,
mondalPRB12,matanoJPSJ13}, and the heavy-fermion systems
CePt$_3$Si~\cite{bauerPRL04} and CeIrSi$_3$~\cite{Sugitani06}.
Different types of topological surface states in fully gapped and nodal NCSs
have recently been classified~\cite{schnyder2012}  and their properties have
been investigated
extensively~\cite{schnyder2011,schnyder2012,brydon2011,tanakaPRL2010,iniotakisPRB07}.
It was found that, depending on the crystal point group and the superconducting
pairing symmetry, NCSs can exhibit either dispersing Majorana surface
states, zero-energy surface flat bands, or arc surface
states, as well as edge modes which are not topologically protected.
Remarkably, these surface states generally exhibit an intricate helical
spin texture. That is, due to spin-orbit interactions, the spin orientation of
the quasiparticle surface states is coupled to their
momentum~\cite{vorontsovPRL08,luYip10,brydonNJP2013,schnyderPRL2013,
hofmannprb2013}.

The nontrivial spin texture of NCS surface states is known to have
important consequences for the surface
physics~\cite{yadaPRB2011,matsuuraNJP2013,vorontsovPRL08,luYip10,brydonNJP2013,schnyderPRL2013,hofmannprb2013,wongPRB2013}. For
example, the helical character of the spin texture forbids spin-independent
scattering between states on opposite sides of the surface Brillouin
zone~\cite{schnyderPRL2013}, which leaves signatures in
Fourier-transformed scanning tunneling
spectroscopy~\cite{hofmannprb2013}. The spin polarization of the edge
states also determines their coupling to magnetic exchange
fields~\cite{yadaPRB2011,matsuuraNJP2013,brydonNJP2013,schnyderPRL2013,wongPRB2013}. Most
strikingly, this is responsible for the appearance of a strong
interface current in heterostructures involving a nodal NCS and a
ferromagnet, which can be used to deduce the existence of
nondegenerate zero-energy flat bands in the
NCS~\cite{brydonNJP2013,schnyderPRL2013}. Furthermore, the nonzero
surface-state spin polarization also gives rise to surface spin
currents~\cite{vorontsovPRL08,luYip10}. Despite the major role played by
the helical spin texture of the surface states in the physics of NCSs, 
it has not yet been systematically studied. In particular, in order to
exploit it as a test of the pairing symmetry of NCSs, and
hence their topological properties, it is essential to know how the
helical spin texture depends on the key variables, namely the spin-orbit
coupling and the ratio of singlet to triplet gaps.

In this paper, we use a quasiclassical theory to investigate the spin
character of topological surface states in both fully gapped and nodal
NCSs and its dependence upon  spin-orbit coupling and
superconducting pairing symmetry. In particular, in the case of nodal
NCSs it is shown that the helical spin texture of the surface states is
controlled by the spin polarization of the  bulk states at the
gap nodes, and thus by the spin structure in the \emph{normal} state.
In our calculation we focus mainly upon two complementary models of
two-dimensional NCSs, but we also survey three-dimensional models of
direct relevance to experimental systems. In the second part of the
paper, we show how the existence of the
spin polarization can be evidenced by tunneling into the NCS through a
ferromagnetic insulator. Specifically, we show that the zero-bias
conductance is very sensitive to the orientation of the barrier
magnetization, and also contains signatures of the pairing symmetry.

\section{Model Hamiltonian and symmetries}~\label{model}

We study subgap states localized at the edge or surface of
NCSs described by the Bogoliubov-de Gennes (BdG) Hamiltonian
\begin{eqnarray}  \label{BdgEQ}
\check{H} ( {\bf k} )
=
\left(\begin{array}{cc}
\hat{h}_{0} & \hat{\Delta} \\
\hat{\Delta}^{\dagger} & -(\hat{h}_{0})^{\ast}
\end{array}\right) .
\end{eqnarray}
Here, $\hat{h}_{0}$ describes the normal part of the Hamiltonian,
\beq
\hat{h}_{0} = \left( \frac{\hbar^2}{2m}\, {\bf k}^2 -
\mu\right)\hat{\sigma}^{0} + \lambda  \, {\bf l}_{\bf k}
\cdot\hat{\pmb{\sigma}} \label{eq:ncsh0} ,
\eeq
where $m$ is the effective mass, $\mu$ the chemical potential,
$\lambda$ the spin-orbit coupling strength, $\hat{\sigma}^{0}$  the $2
\times 2$ unit matrix, $\hat{\bm \sigma}$ the vector of Pauli
matrices, and ${\bf l}_{\bf k}$ the antisymmetric (i.e., odd in
$\mathbf{k}$) spin-orbit coupling pseudovector. The
Hamiltonian $\hat{h}_{0}$  in \eq{eq:ncsh0} is diagonalized in the
helicity basis,
$\hat{h}_0 = \mbox{diag}(\xi^{+}_{\bf k},\xi^{-}_{\bf k})$, where
\beq
\xi^{\pm}_{{\bf k}} = \left( \frac{\hbar^2}{2m} {\bf k}^2 -
\mu\right) \pm \lambda|{\bf l}_{\bf k}|\,
\eeq
are the dispersions of the positive ($+$) and negative ($-$) helicity
bands.

Due to the breaking of inversion symmetry, the superconducting gap function
\beqarray  \label{eq:ncspair}
\hat{\Delta}
& = &
 \left[ \psi_{\bf k} \hat{\sigma}^{0} +  {\bf d}_{\bf k}
   \cdot\hat{\pmb{\sigma}}\right]  i\hat{\sigma}^{y}
\eeqarray
generically contains both a spin-singlet component
$\psi_{\bf k} = \Delta_s f ( \widetilde{\bf k} )= q
\Delta_0  f ( \widetilde{\bf k} )$ and a spin-triplet component ${\bf
  d}_{\bf k} = \Delta_{t} f (\widetilde{\bf k}) \, {\bf l}_{\widetilde{\bf
  k}} = (1 - q)\Delta_0 f(\widetilde{\bf k})\,{\bf l}_{\widetilde{\bf
  k}}$~\cite{frigeriPRL04}, where $q$ tunes the system from
purely triplet ($q=0$) to purely singlet ($q=1$) pairing. We also
introduce the dimensionless momentum $\widetilde{\bf k} = {\bf
  k}/k_F$, where $k_{F} =
(2m\mu)^{1/2}/\hbar$ is the Fermi wavevector in the absence of the
spin-orbit coupling.
The orientation of the vector $\mathbf{d}_\mathbf{k}$ parallel
to ${\bf l}_{\bf k}$ implies pairing only between states on the same
helicity Fermi surface, opening the gaps
\beq \label{deltaf}
\Delta^\pm_{\bf k} = \big[q \pm (1-q)|{\bf l}_{\widetilde{\bf k}}|\big]\,
  \Delta_0 f(\widetilde{\bf k}) .
\eeq
The form factor $f ({\bf k})$ determines the
orbital-angular-momentum pairing state. In the following we focus on
two cases:   $f ( {\bf k })=1$ for a NCS with ($s + p$)-wave pairing
symmetry~\cite{satoFujimoto2009,tanakaNagaosaPRB09,iniotakisPRB07,vorontsovPRL08,luYip10}
and $f ( {\bf k }) = 2 k_x  k_y $ for a
($d_{xy}+p$)-wave pairing state~\cite{tanakaPRL2010,schnyderPRL2013}.

The momentum dependence of the spin-orbit pseudovector ${\bf l}_{\bf
  k}$ is restricted by the symmetries of the noncentrosymmetric
crystal. We consider
three different crystallographic point groups: tetragonal
$C_{4v}$, cubic $O$, and monoclinic $C_2$.
Within a small-momentum expansion around the $\Gamma$
point~\cite{Samokhin2009}, the  vector $\mathbf{l}_\mathbf{k}$
for the tetragonal point group $C_{4v}$ is written as
\begin{eqnarray}
\fl \qquad {\bf l}_{\bf k} = \hat{\bf x}\,   k_y
- \hat{\bf y}\,   k_x . \label{eq:Rashba}
\end{eqnarray}
Examples of $C_{4v}$ NCSs are CePt$_3$Si~\cite{bauerPRL04} and
CeIrSi$_3$~\cite{Sugitani06}. This form of $\mathbf{l}_\mathbf{k}$
is often referred to
as Rashba spin-orbit coupling. For the cubic point group $O$ we have
\begin{eqnarray}
\fl \qquad {\bf l}_{\bf k} =   \hat{\bf x}\,   k_x (1 + g_2 [ k_y^2 +
  k_z^2] )+  \hat{\bf y}\,   k_y  (1+ g_2 [ k_x^2 + k_z^2  ])
+ \hat{\bf z}\,   k_z  ( 1 + g_2 [ k_x^2 + k_y^2 ]),
\end{eqnarray}
where we include the second-order spin-orbit coupling $g_2$. This  point
group is
relevant for Li$_2$Pd$_x$Pt$_{3-x}$B~\cite{yuanPRL06,nishiyama07} and
Mo$_3$Al$_2$C~\cite{bauerPRB10,karki10}. For the monoclinic group
$C_2$, which is relevant for BiPd
\cite{joshiPRB11,mondalPRB12,matanoJPSJ13}, we  have
\begin{eqnarray} \label{eq:Lvector}
\fl \qquad {\bf l}_{\bf k}
= \hat{\bf x}\, ( a_1 k_x + a_2 k_y)
+ \hat{\bf y}\, ( a_3 k_x + a_4 k_y)
+ \hat{\bf z}\, a_5 k_z .
\end{eqnarray}
A three-dimensional $C_2$ NCS with $\mathbf{l}_\mathbf{k}$ given
by Eq.~\eqref{eq:Lvector} generically exhibits
nodal rings in the BdG spectrum. The number of these nodal rings and
their position in the Brillouin zone depend on the particular values
of the parameters $a_i$ and the singlet-triplet ratio $q$. For the
numerical calculations, we set $a_i = 1$ for all
$i=1, \ldots, 5$. Other parameter choices  give qualitatively similar results.

The BdG Hamiltonian $\check{H}({\bf k})$ possesses all three symmetries which
form the basis of the topological ten-fold way classification
\cite{schnyderPRB08,kitaev22,schnyderAIP}: time-reversal, particle-hole, and
chiral symmetry. Time-reversal acts as
$\check{U}_T^{\dag} \check{H} ( {\bf k} ) \check{U}_T =
\check{H}^{\!\mathrm{T}} ( - {\bf k} ) $,
where  $\check{U}_T = \tau^0 \otimes i  \sigma^y$ and $\tau^i$ are the
Pauli matrices in Nambu space.
The time-reversal operator squares to $\check{U}_T \check{U}_T^\ast =
  (\tau^0 \otimes i\sigma^y)(\tau^0 \otimes i\sigma^y)
  = \tau^0 \otimes (-\sigma^0) = - \mathbbm{1}$,
where $\mathbbm{1}$ is the $4\times 4$ unit matrix.
Particle-hole symmetry acts on
$\check{H} ( {\bf k} )$ as  $\check{U}_C^{\dag} \check{H} ( {\bf k} )
\check{U}_C =  -\check{H}^{\!\mathrm{T}} ( - {\bf k} )$,
where  $\check{U}_C =  \tau^x \otimes   \sigma^0 $.
The particle-hole-conjugation operator squares to
$\check{U}_C \check{U}_C^\ast = (\tau^x \otimes \sigma^0)
  (\tau^x \otimes \sigma^0) = \tau^0 \otimes \sigma^0 =  + \mathbbm{1}$.
Hence, $\check{H} ({\bf k} )$ belongs to symmetry class DIII.
Combining time-reversal and particle-hole symmetry yields the
so-called chiral symmetry, which acts on the BdG Hamiltonian
as $\check{U}_S  \check{H} ( {\bf k} ) + \check{H} (  {\bf k} )
\check{U}_S = 0 $, where $\check{U}_S = i \check{U}_T \check{U}_C = - \tau^x
\otimes \sigma^y$.

\subsection{Surface states}

In order to obtain the surface-state wavefunctions we solve the
BdG equations
\beq
\left(\begin{array}{cc}
\hat{h}_{0} & \hat{\Delta} \\
\hat{\Delta}^{\dagger} & -(\hat{h}_{0})^{\ast}
\end{array}\right)\Psi({\bf r}) = E\Psi({\bf r}) \label{eq:BdGeq}
\eeq
subject to the boundary conditions
$\left. \Psi({\bf r})\right|_{r_\perp=0}=0$ and $\left. \Psi({\bf
  r})\right|_{r_\perp \to + \infty}=0$, where $r_\perp$ is the
coordinate normal to the surface. The wavevector component ${\bf
  k}_\parallel$ parallel to the surface is a good quantum number by
translational invariance, and so we henceforth work with the
Fourier-transformed wavefunction $\Psi({\bf k}_\parallel;r_\perp)$,
which is an eigenfunction of the one-dimensional Hamiltonian
$\check{H}({\bf k}_\parallel)$.  Depending on the value of ${\bf k}_\parallel$,
we distinguish
 two cases when constructing the wavefunction ansatz: the
surface momentum lies within both projected
Fermi surfaces or only within the projected (larger)
negative-helicity Fermi surface.

\subsubsection{Surface momentum within projection of both Fermi surfaces.}

In this case there are wavevectors on both positive and negative
helicity Fermi
surfaces which project onto the surface momentum ${\bf k}_\parallel$,
specifically ${\bf k}^{}_\pm=({\bf k}_\parallel,k^{}_{\perp,\pm})$ and
${\bf k}^\prime_\pm=({\bf
  k}_\parallel,{k}^{\prime}_{\perp,\pm})$, where the perpendicular
wavevector components $k^{}_{\perp,\pm}$ and ${k}^\prime_{\perp,\pm}$
have opposite sign. The
wavefunction ansatz for the bound state is then a superposition of
evanescent states in the various channels,
\beqarray
\Psi({\bf k}_\parallel; r_\perp) & = & \sum_{\nu = \pm}
  \sum_{{\bf k}={\bf k}^{}_\nu,{\bf k}^\prime_\nu}\,
  \alpha_{\nu}({\bf k})\psi_{\nu}( {\bf k})\,
  e^{ik_{\perp} r_\perp}e^{-\kappa^{\nu}_{{\bf k}} r_\perp}, \label{eq:Psicase1}
\eeqarray
where the spinors in~\eq{eq:Psicase1} are given by
\beq
\psi_{\pm}({\bf k}) = \left(\begin{array}{cccc}1, & \pm \frac{l^{x}_{\bf k} +
    il^{y}_{\bf k}}{|{\bf l}_{\bf k}| \pm l^{z}_{\bf k}}, & \mp\frac{l^{x}_{\bf k}
    + il^{y}_{\bf k}}{|{\bf l}_{\bf k}| \pm l^{z}_{\bf k}}\, \gamma^{\pm}_{\bf k}, &
  \gamma^{\pm}_{\bf k}\end{array}\right)^{ \mathrm{T}} , \label{eq:spinorcase1}
\eeq
with
\begin{eqnarray}
\gamma^{\pm}_{\bf k}
&=&
\frac{1}{\Delta^\pm_{\bf k}}\left[E -
  i \sgn(v^{\pm}_{F,\perp}({\bf k}))\sqrt{|\Delta^\pm_{\bf k}|^2 - E^2}\right] ,
   \\
\kappa^{\pm}_{\bf k}
&=&
\frac{1}{\hbar\,|v^{\pm}_{F,\perp}({\bf k})|}\sqrt{|\Delta^{\pm}_{\bf
    k}|^2-E^2}, \label{eq:kappapm}
\end{eqnarray}
and $v^{\pm}_{F,\perp}({\bf k})$ is the component of the Fermi
velocity normal to the surface.
A bound state is realized when it is possible to choose nonzero
coefficients $\alpha_{\nu}({\bf k})$ in~\eq{eq:Psicase1} such that the
wavefunction obeys the normalization condition
\beq
1 = \int^{\infty}_{0} d r_\perp \Psi^\dagger({\bf k}_\parallel; r_\perp)\,
  \Psi({\bf k}_\parallel; r_\perp)
\eeq
and vanishes at the surface.  The former condition is satisfied if
$|E|<\min\{|\Delta^{\pm}_{{\bf
  k}^{}_\pm}|,|\Delta^{\pm}_{{\bf k}^{\prime}_\pm}|\}$.
From \eq{eq:Psicase1} we see that the latter condition is equivalent to
\beq
 \det\left[\begin{array}{cccc}
\psi_+({\bf k}^{}_+) & \psi_+({\bf k}^\prime_+) & \psi_-({\bf
  k}^{}_-) & \psi_-({\bf k}^\prime_-)\end{array}\right] = 0.
\label{eq:cond2prop}
\eeq
Solutions of this equation satisfying
$|E|<\min\{|\Delta^{\pm}_{{\bf
  k}^{}_\pm}|,|\Delta^{\pm}_{{\bf k}^{\prime}_\pm}|\}$ are
the bound-state energies, which can  belong to either dispersing or
zero-energy flat bands.

\subsubsection{Surface momentum only within projection of negative-helicity
Fermi surface.}

In the case that there are propagating solutions only on the
negative-helicity Fermi surface, the positive-helicity
components of the wavefunction ansatz in~\eq{eq:Psicase1} are replaced by
\beq
 \big[ \alpha_{e,+}({\bf p})\phi_e({\bf
  p}) + \alpha_{h,+}({\bf p})\phi_h({\bf
  p}) \big] \, e^{ip_{\perp} r_\perp} , \label{eq:Psicase2}
\eeq
where ${\bf p} = ({\bf k}_\parallel,p_\perp)$ satisfies $\xi^{+}_{\bf
  p} = 0$ and the imaginary part of $p_\perp$ is positive. The spinors
$\phi_e({\bf p})$ and $\phi_h({\bf p})$ describe an electronlike or holelike
state in the absence of the pairing potential,
\beqarray
\phi_e({\bf p}) & = & \left(\begin{array}{cccc}1, &
  \frac{l^{x}_{\bf p}+il^{y}_{\bf p}}{|{\bf l}_{\bf p}| + l^{z}_{\bf p}}, &
  0, & 0\end{array}\right)^{\!\mathrm{T}}, \\
\phi_h({\bf p}) & = & \left(\begin{array}{cccc}0, &
  0, &
  -\frac{l^{x}_{\bf p}+il^{y}_{\bf p}}{|{\bf l}_{\bf p}| + l^{z}_{\bf
      p}}, & 1\end{array}\right)^{\!\mathrm{T}} .
\eeqarray
The condition for the existence of the bound state now becomes
 \beq
\det\left[\begin{array}{cccc}
\phi_e({\bf p}) & \phi_h({\bf p}) & \psi_-({\bf
  k}^{}_-) & \psi_-({\bf k}^\prime_-)\end{array}\right] = 0.
\label{eq:condxprop}
\eeq
Unlike~\eq{eq:cond2prop}, this only allows for the existence of
\emph{nondegenerate} zero-energy flat bands, which occur whenever
$\sgn (\Delta^{-}_{{\bf k}^{}_-}) = -\sgn (\Delta^{-}_{{\bf k}^\prime_-})$.

\subsubsection{Symmetries of the wavefunctions.}

The symmetries characterizing the bulk BdG Hamiltonian remain valid
for the edge states. Hence, for every
surface-state wavefunction $ \Psi ( {\bf k}_{\parallel};r_\perp)$
satisfying  $\check{H} ( {\bf k}_{\parallel}  ) \Psi ( {\bf
k}_{\parallel};r_\perp)
= E ( {\bf k}_{\parallel} ) \Psi ( {\bf k}_{\parallel};r_\perp) $, there is
a time-reversed partner
 $\check{U}_T   \Psi^{\ast}  ( {\bf k}_{\parallel};r_\perp)$, which is an
eigenfunction of $\check{H} (- {\bf k}_{\parallel}  )$ with the same
energy
 $ E (- {\bf k}_{\parallel} ) =  E ( {\bf k}_{\parallel} )$, i.e.,
\beq \label{TRSeins}
\check{H} (- {\bf k}_{\parallel}  )
 \check{U}_T   \Psi^{\ast}  ( {\bf k}_{\parallel} ;r_\perp)
 =
 E ( {\bf k}_{\parallel} )
 \check{U}_T   \Psi^{\ast}  ( {\bf k}_{\parallel} ;r_\perp)   .
\eeq
Due to Kramer's theorem, $\Psi ( {\bf k}_{\parallel};r_\perp)$ and
$\check{U}_T   \Psi^{\ast}  ( {\bf k}_{\parallel};r_\perp)$ are orthogonal
for all ${\bf k_{\parallel} }$.
Similarly, particle-hole symmetry dictates that for
every surface-state eigenfunction $\Psi ( {\bf k}_{\parallel};r_\perp)$ there is
a particle-hole-reversed partner
$\check{U}_C \Psi^{\ast} ( {\bf k}_{\parallel};r_\perp)$, which is an
eigenfunction of $\check{H} ( - {\bf k}_{\parallel} )$ with energy $- E ( {\bf
  k}_{\parallel} )$, i.e.,
\beq \label{PHSeins}
\check{H} (- {\bf k}_{\parallel}  )
 \check{U}_C   \Psi^{\ast}  ( {\bf k}_{\parallel};r_\perp)
 =
- E ( {\bf k}_{\parallel} )
 \check{U}_C   \Psi^{\ast}  ( {\bf k}_{\parallel};r_\perp)   .
\eeq
Finally, the presence of chiral symmetry requires that for every
surface state  $\Psi ( {\bf k}_{\parallel};r_\perp) $ with energy
$E ( {\bf k}_{\parallel} )$ there is
a chiral-symmetric partner $\check{U}_S   \Psi  ( {\bf k}_{\parallel}
;r_\perp)$ with energy $-E ( {\bf k}_{\parallel} )$, i.e.,
\beq \label{SLSeins}
\check{H} (  {\bf k}_{\parallel}  )
 \check{U}_S   \Psi   ( {\bf k}_{\parallel};r_\perp)
 =
- E ( {\bf k}_{\parallel} )
 \check{U}_S   \Psi  ( {\bf k}_{\parallel};r_\perp)   .
\eeq
We observe that all eigenfunctions of $\check{H} (  {\bf k}_{\parallel}  )$ can
be chosen to be simultaneous  eigenfunctions of  $\check{U}_S$ with
chirality eigenvalue $\pm 1$ \cite{satoPRB2011}.

\subsection{Spin polarization}

We define the  $\mu$-component of the spin polarization of the
surface state with energy $E$ and surface momentum ${\bf k}_\parallel$ as the
expectation value
\beq
\rho^{\mu}_{\mathrm{tot}} ( E, {\bf k}_{\parallel}
) = \int_{0}^{\infty} dr_\perp \Psi^{\dagger}({\bf
  k}_\parallel;r_\perp) \check{S}^{\mu} \Psi({\bf
  k}_\parallel;r_\perp)
\eeq
of the total spin operator $\check{S}^{\mu}$ with respect to the
 wavefunction $\Psi({\bf k}_\parallel;r_\perp)$. The total spin operator
in Nambu space reads
\beq
\check{S}^{\mu}
= \left(
\begin{array}{c c}
\sigma^{\mu} & 0 \cr
0 & - \left[ \sigma^{\mu} \right]^{\ast}  \cr
\end{array} \right) ,
\eeq
with  $\mu = x,y,z$. Note that the coupling of the surface states
to an external exchange field is determined by the total spin
polarization~\cite{brydonNJP2013,schnyderPRL2013}. On the other hand, the surface spin current of
NCSs can be
understood in terms of the spin polarization of the electronlike (or
holelike) part of the surface-state wavefunction $\Psi  ( {\bf
  k}_{\parallel};r_\perp) $~\cite{luYip10,vorontsovPRL08}. Hence, it is  useful to define
an electronlike (holelike) spin polarization
\beq
\rho^{\mu}_{\mathrm{e(h)}} ( E, {\bf k}_{\parallel}
) = \int_{0}^{\infty} dr_\perp \Psi^{\dagger}({\bf
  k}_\parallel;r_\perp) \check{S}^{\mu}_{e(h)} \Psi({\bf
  k}_\parallel;r_\perp)
\eeq
in terms of the electronlike and holelike spin operators
\beq
\check{S}^{\mu}_{\mathrm{e}} = \left(\begin{array}{cc}
  \sigma^\mu & 0 \cr
  0 & 0 \cr
  \end{array}\right), \qquad
\check{S}^{\mu}_{\mathrm{h}} = \left(\begin{array}{cc}
  0 & 0 \cr
  0 & - \left[ \sigma^\mu \right]^\ast \cr
  \end{array}\right) ,
\eeq
respectively.

The symmetry properties of the edge-state wavefunctions are reflected in
their spin polarization.  Specifically, the various symmetries give the
following constraints:
\begin{itemize}
\item time-reversal symmetry:
\begin{eqnarray}
\fl \quad
\rho^{\mu}_{\mathrm{e(h)}} ( E, {\bf k}_{\parallel} ) =  - \rho^{\mu}_{\mathrm{e (h)} } ( E, - {\bf k}_{\parallel} ) ,
\quad
\rho^{\mu}_{\mathrm{tot}} ( E, {\bf k}_{\parallel} )
= - \rho^{\mu}_{\mathrm{tot}} ( E, - {\bf k}_{\parallel} ) ,
\end{eqnarray}
\item particle-hole symmetry:
\begin{eqnarray}
\fl \quad
\rho^{\mu}_{\mathrm{e(h)}} ( E, {\bf k}_{\parallel} ) =  - \rho^{\mu}_{\mathrm{h (e)} } ( -  E, - {\bf k}_{\parallel} ) ,
\quad
\rho^{\mu}_{\mathrm{tot}} ( E, {\bf k}_{\parallel} )
= - \rho^{\mu}_{\mathrm{tot}} ( - E, -{\bf k}_{\parallel}) ,
\end{eqnarray}
\item chiral symmetry:
\begin{eqnarray}
\fl \quad
\rho^{\mu}_{\mathrm{\mathrm{e(h)}}} ( E, {\bf k}_{\parallel} ) = \rho^{\mu}_{\mathrm{h (e)} } ( - E,  {\bf k}_{\parallel} ) ,
\quad
\rho^{\mu}_{\mathrm{tot}} ( E, {\bf k}_{\parallel})
=
\rho^{\mu}_{\mathrm{tot}} ( - E, {\bf k}_{\parallel}) .
\end{eqnarray}
\end{itemize}
Due to the chiral and particle-hole symmetries, it is only necessary
to consider the total spin polarization for the bound states with
nonnegative  energies. Time-reversal symmetry requires that the spin
polarization is an odd function of the surface momentum, and so
there will be no spin accumulation at the surface, although a surface spin
current is permitted~\cite{vorontsovPRL08,luYip10}.

In the following we present results only for the total spin
polarization and we thus drop the subscript ``tot''.
To evaluate the spin polarization, it is necessary to determine the
coefficients $\alpha_{\nu}({\bf k})$ in the wavefunctions. This is
equivalent to determining the null space of the matrices with column
vectors given by the spinors in the wavefunction ansatz. In general it
is necessary to numerically calculate the coefficients and the spin
polarization.
In our numerical calculations we take the BCS correlation length
$\xi_0 = 2\hbar v_F/\pi\Delta_0 = 100\, k_F^{-1}$ where $v_F = \hbar
k_F/m$;
although this is at the lower limit of physical values, larger values
only result in minor quantitative changes. We also introduce the
dimensionless spin-orbit coupling $\widetilde{\lambda} = \lambda
m/\hbar^2k_F$. Due to the symmetries of the spin polarization, we
restrict ourselves to nonnegative bound-state energies and henceforth
drop the energy argument in  the spin  polarization, $\rho^\mu(E,{\bf
  k}_\parallel) \rightarrow \rho^\mu({\bf k}_\parallel)$.

\section{Edge states of two-dimensional NCSs}

We commence by considering the $(10)$ edge states of two-dimensional NCSs with
$C_{4v}$ point group, which can be obtained by restricting the three-dimensional
model to the $k_z=0$ plane. The normal-state Fermi surface
consists of two concentric circles with radii
$k_{F,\pm} = k_{F}\,[(1 + \widetilde{\lambda}^2)^{1/2} \mp
\widetilde{\lambda}]$.
The two choices for the superconducting form factor $f({\bf
k})$, cf.~Eq.~\eqref{deltaf}, give qualitatively different topologies and
thus very different edge states. The system is fully gapped in the ($s+p$)-wave
case for all values of the singlet-triplet parameter $q$, except at $q=q_c =
k_{F,-}/(k_F+k_{F,-})$ where the negative-helicity gap vanishes. This
marks the boundary between the topologically nontrivial ($q<q_c$) and trivial
($q>q_c$) regimes, and the topology is characterized by a $\mathbb{Z}_2$ bulk
topological invariant. In agreement with the bulk-boundary correspondence,
helical edge states with Majorana zero-energy modes are present only in the
topological state. In contrast, a bulk topological invariant cannot be defined
for the nodal ($d_{xy}+p$)-wave NCS. Nevertheless, this system possesses
topologically protected flat-band zero-energy edge states. The topological
protection arises by interpreting the edge state at edge momentum $k_y$ to be
the edge state of a one-dimensional Hamiltonian $\check{H}(k_y)$ which falls
into class AIII. The topology of this Hamiltonian is characterized by a
$\mathbb{Z}$ number; in particular, when this number evaluates
to $\pm1$, the edge states are 
\emph{nondegenerate}, i.e., they have a Majorana
character~\cite{yadaPRB2011,schnyder2011,schnyder2012,brydon2011,tanakaPRL2010,dahlhaus2012}.

\subsection{($s+p$)-wave NCS}

\begin{figure}[t]
\includegraphics[clip,width=\columnwidth]{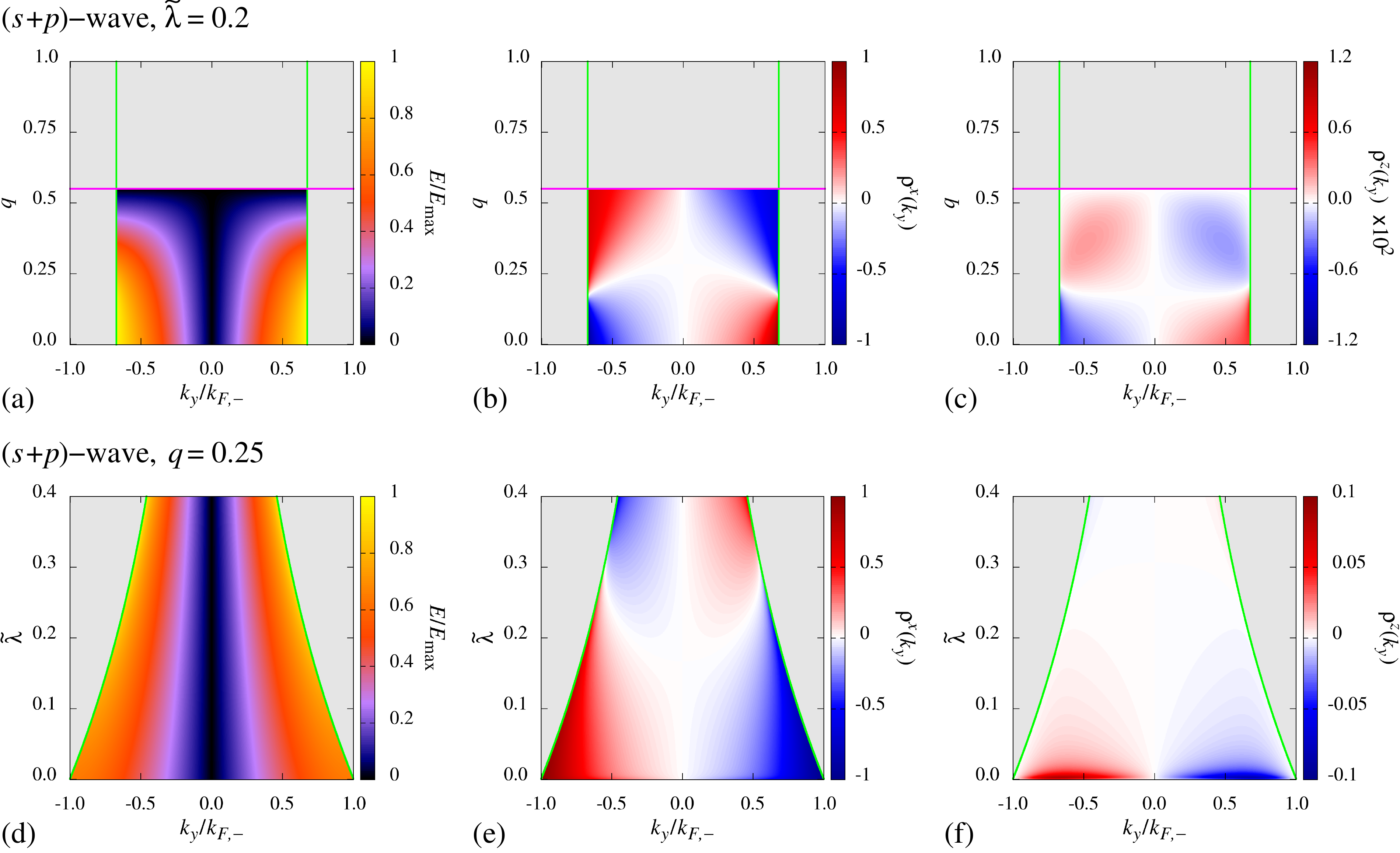}%
\caption{\label{2DC4vsp}
Evolution of the momentum-resolved edge-state spectrum and spin
polarization at the $(10)$ edge of the 2D $C_{4v}$ ($s$+$p$)-wave NCS
 (a)--(c)  as a function of  the singlet-triplet parameter $q$ and
 (d)--(f) as a function of the
spin-orbit strength $\widetilde{\lambda}$. The first column gives the
dispersion
of the nonnegative-energy edge states compared to the maximum
energy in the plot, $E_{\rm{max}}$, while the second and third columns
give the total $x$- and $z$-spin polarization, respectively. The
$y$-spin polarization  vanishes. Grey regions indicate the absence
of any edge state. The  green  lines indicate the
projected edge of the positive-helicity Fermi surface, while the
horizontal magenta line in panels (a)--(c) indicates the
negative-helicity gap closing at $q=q_c$.}
\end{figure}

In figure~\ref{2DC4vsp} we plot the dispersion  and the spin
polarization  $\rho^{\mu}(k_y)$  of the  edge states with nonnegative
energy in the
($s+p$)-wave phase. As can be seen from panels (a) and (d), the helical
edge states are only present in the topologically nontrivial regime
($q<q_c$) and within the projection of the positive-helicity
Fermi surface ($|k_{y}| \leq k_{F,+}$).
The remaining panels of figure~\ref{2DC4vsp} reveal that the edge states
exhibit a spin polarization in the  $xz$  plane, with
a particularly strong component along the $x$ axis. The spin
polarization depends upon the singlet-triplet parameter $q$ and the
spin-orbit coupling strength $\widetilde{\lambda}$, and changes sign
as these quantities are increased.
Note that the spin polarization is not determined by the
topological properties of the system  alone: in the topologically
nontrivial
state it is possible to continuously deform the system to a helical
$p$-wave superconductor without spin-orbit coupling (i.e., $\Delta_s=0$ and
$\widetilde{\lambda}=0$), for which the edge states have vanishing
spin polarization.

The dramatic variation of the spin polarization is controlled by the
spin-orbit coupling and the gap structure. Focusing upon the $x$-spin
polarization, we gain insight into their interplay by first considering
the polarization close to $|k_y|=k_{F,+}$, where the subgap states
enter the continuum at energy $E=\min\{\Delta^{+}_{\bf
  k},|\Delta^{-}_{\bf k}|\}$. As $|k_y|$ approaches $k_{F,+}$, the edge
states smoothly evolve to match the bulk wavefunctions at the edge of the
continuum. The $x$-spin polarization of the edge state will
hence also evolve to match that of the continuum states with
transverse momentum $k_y$, which for the
$\nu=\pm$ helicity band is given by $\nu k_y/k_{F,\nu}$. Thus, when
the negative-helicity gap is the smallest (i.e., for 
 $q>\widetilde{\lambda}/(1+{\widetilde{\lambda}})$), the edge
states  close  to
the gap edge are dominated by the negative-helicity components, and
hence have spin polarization $-\sgn(k_y)\,
k_{F,+}/k_{F,-}$. On the other hand, the edge states close to the
continuum have spin polarization $\sgn(k_y)$ when the positive-helicity
gap is the smallest (i.e., for
 $q < \widetilde{\lambda}/(1+{\widetilde{\lambda}})$). This is in
excellent agreement with the 
numerical results. This argument also holds away from
the gap edges: the full results for the $x$-spin polarization
is well represented by
\beq
\rho^{x}(k_y) \approx \sum_{\nu = \pm}\nu\,
  \frac{k_y}{k_{F,\nu}} \int^{\infty}_{0} dx\, [{\cal
    P}_{\nu}\Psi(k_y;x)]^\dagger\, {\cal P}_{\nu}\Psi(k_y;x) ,
\eeq
where ${\cal P}_{\nu}$ projects onto the $\nu$ helicity
components in~\eq{eq:Psicase1}. That is, the variation of the $x$-spin
polarization reflects the relative strength of the positive- and
negative-helicity components of the wavefunction.

\begin{figure}[t]
\includegraphics[clip,width=\columnwidth]{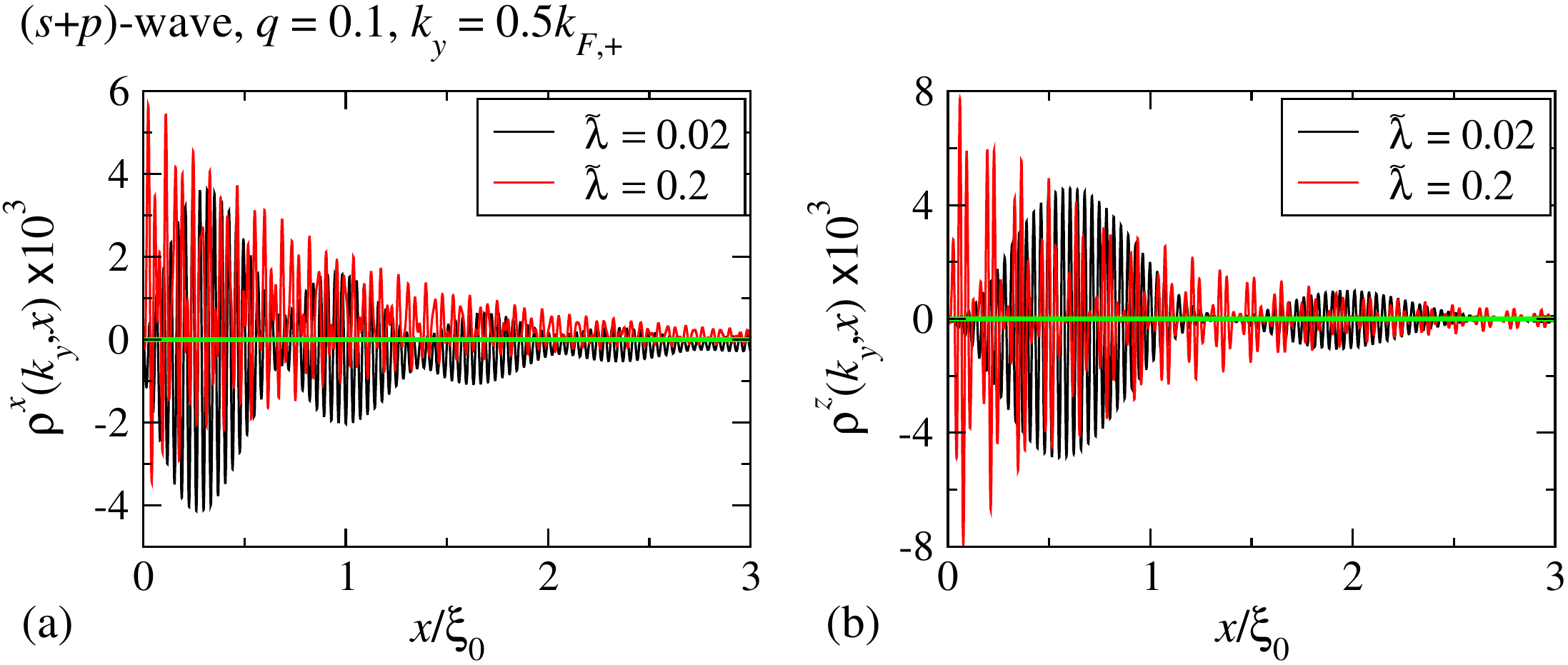}%
\caption{\label{spindensity}Typical plots of the spin  density
$\rho^\mu(k_y,x)$ of an edge
state of the ($s+p$)-wave NCS for (a) $\mu=x$ and (b) $\mu=z$. We take
$q=0.1$ and $k_y=0.5\,k_{F,+}$. The green  line indicates
zero. }
\end{figure}

Such an argument cannot be made for the $z$-spin polarization, 
however, as the
bulk states of the  two-dimensional NCS are polarized in the  $xy$
plane. This also holds for the spinors  in~\eq{eq:spinorcase1} comprising
the wavefunction, i.e., $\psi^\dagger_{\pm}({\bf
  k})\check{S}^z\psi_{\pm}({\bf k}) = 0$. The $z$-spin polarization
thus arises entirely due to the interference between the different
channels in the wavefunction ansatz  in~\eq{eq:Psicase1}; this is in
contrast to the $x$-spin polarization, where $\psi^\dagger_{\pm}({\bf
  k})\check{S}^x\psi_{\pm}({\bf k})$ is generally nonzero.
As such, the spin density
\beq
\rho^{\mu}(k_y,x) = \Psi^\dagger(k_y;x)\check{S}^\mu\Psi(k_y;x)
\eeq
shows damped oscillations about zero for $\mu=z$, whereas for $\mu=x$
 it oscillates about a finite value. It
hence follows that the integrated $z$-spin density will
be much smaller than that for the $x$-spin density, in agreement with
the numerics.  For illustration, we plot in figures~\ref{spindensity}(a) and
(b) typical examples of the $x$- and $z$-spin densities, respectively.

\subsection{($d_{xy}+p$)-wave NCS}

\begin{figure}[t]
\includegraphics[clip,width=\columnwidth]{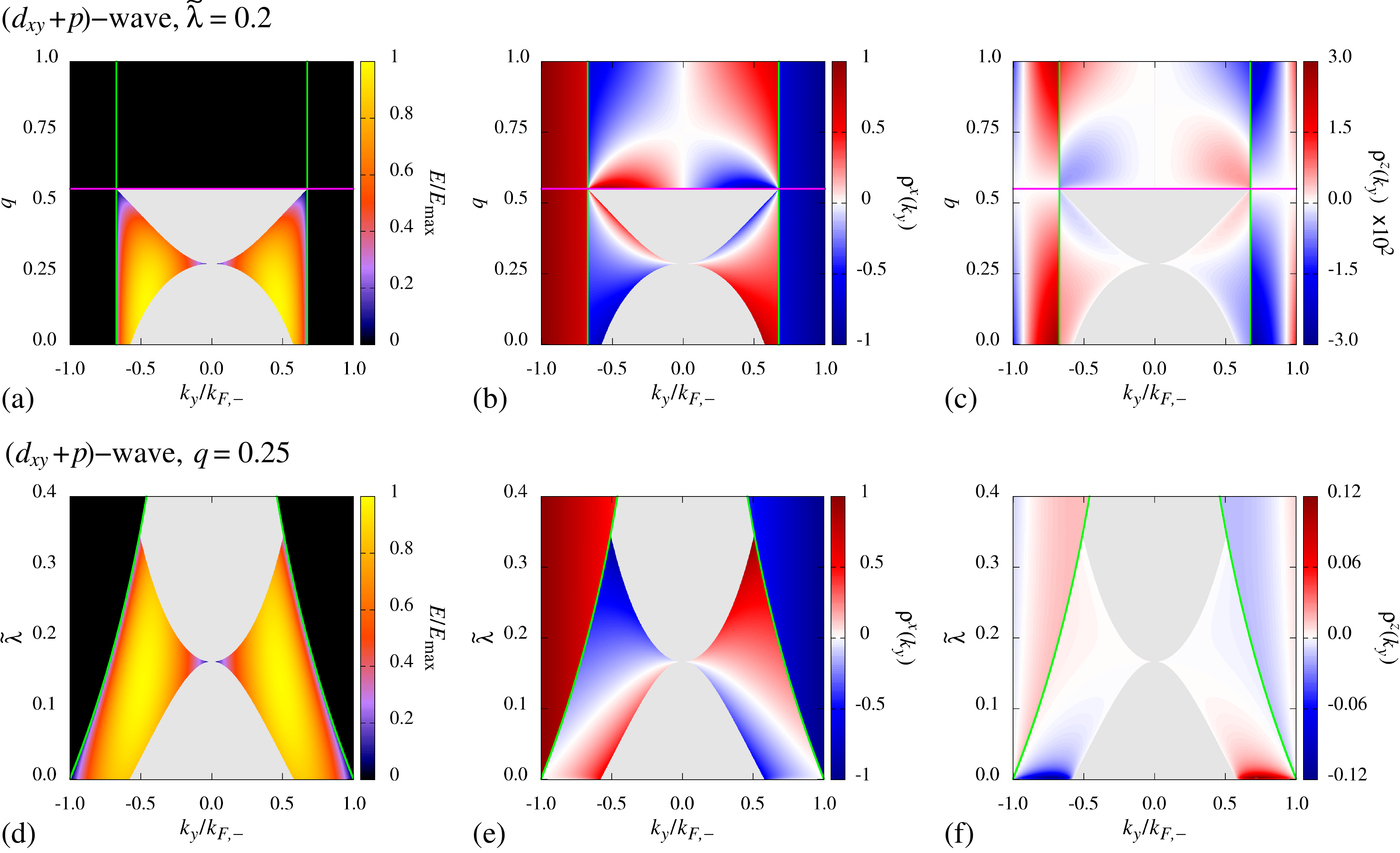}%
\caption{\label{2DC4vdp}
 Evolution of the momentum-resolved edge-state spectrum and spin
polarization for the 2D $C_{4v}$ ($d_{xy}$+$p$)-wave NCS
(a)--(c) as a function of the singlet-triplet parameter $q$ and
(d)--(f) as a function of the
spin-orbit strength $\widetilde{\lambda}$. The first column gives the
 dispersion
of the nonnegative-energy edge states, while the second and third columns
give the total $x$- and $z$-spin polarization, respectively. The
$y$-spin polarization vanishes. Grey regions indicate the absence
of any edge state. Black regions in panels (a), (d) denote flat
zero-energy bands. The green lines indicate the
projected edge of the positive-helicity Fermi surface, while the
horizontal magenta line in panels (a)--(c) indicates the
negative-helicity gap closing at $q=q_c$.}
\end{figure}

The  dispersion and the spin polarization of the edge states with
nonnegative energy in the ($d_{xy}+p$)-wave NCS are shown in
figure~\ref{2DC4vdp}. Zero-energy flat bands are clearly present for
$k_{F,+}<|k_y|< k_{F,-}$ at all values of the singlet-triplet
parameter $q$, and also at $|k_y|<k_{F,+}$ for $q>q_c$. Whereas the
former are nondegenerate, the latter are doubly degenerate, similar to
the flat bands at the edge of a $d_{xy}$-wave superconductor without
spin-orbit coupling.
Dispersing states are present at $|k_y|<k_{F,+}$ for $q<q_c$ and
sufficiently small spin-orbit coupling. Similarly to the
($s+p$)-wave NCS, the edge states are spin-polarized in the
  $xz$ plane, with the
$z$-spin polarization generally much weaker than the $x$-spin
polarization. Note that
the spin polarization of the doubly degenerate zero-energy flat bands
is the sum of the  polarizations of the corresponding two states.

The spin polarization of the dispersing edge states can be
understood by the same arguments as above. More interesting is the
spin polarization of the topologically nontrivial nondegenerate
zero-energy flat bands at $k_{F,+}<|k_y|<k_{F,-}$, in particular their
$x$-spin polarization: as can be seen
in figures~\ref{2DC4vdp}(b) and (e), this  component
shows no dependence upon the
singlet-triplet parameter $q$ or the spin-orbit coupling
$\widetilde{\lambda}$. Furthermore, there appears to be a
discontinuity in the spin polarization across the projected nodal
points of the positive-helicity gap at $k_y = \pm  k_{F,+}$.

It is possible to obtain analytic expressions for the
wavefunctions of the nondegenerate flat-band
states~\cite{tanakaPRL2010,queirozPRB2014}, which read
\beqarray
\fl \Psi(k_y;x) =  N\left\{\left[A_{k_y}\left(e^{ip_+x} -
  e^{-\kappa_-x}\cos(k_{x,-}x)\right) - \frac{k_{F,-} +
    A_{k_y}k_y}{k_{x,-}}\, e^{-\kappa_-x}\sin(k_{x,-}x)\right]\right. \nonumber
\\
 \times \left(\begin{array}{cccc}
i\sgn(k_y[q-(1-q)k_{F,-}]) \,, & 0\,, & 0\,, & 1
\end{array}\right)^{\!\mathrm{T}} \nonumber \\
+ \left[\left(e^{ip_+x} -
  e^{-\kappa_-x}\cos(k_{x,-}x)\right) + \frac{A_{k_y}k_{F,-} +
    k_y}{k_{x,-}}\, e^{-\kappa_-x}\sin(k_{x,-}x)\right] \nonumber
\\
 \left. \times \left(\begin{array}{cccc}
0 \,, & i\sgn(k_y[q-(1-q)k_{F,-}]) \,, & -1\,, & 0
\end{array}\right)^{\!\mathrm{T}}\right\} , \label{eq:flatbandwf}
\eeqarray
where $A_{k_y} = (k_y - p_{+})/k_{F,+}$,
$p_{+} = i(k_y^2 - k_{F,+}^2)^{1/2}$,
 $k_{x,-} = (k_{F,-}^2 - k_y^2)^{1/2}$,
  $N$ is a normalization constant, 
 \beqarray
 \kappa_- =  \frac{4}{\hbar v_F}   \frac{   1 + \frac{k_{F,-}}{k_F}     }
{ 1+  \frac{ k_{F,+}}{k_{F,-}}  } |q - q_c|\Delta_{0} \frac{ k_y}{k_F} ,
\eeqarray
 and the other quantities are as defined in
section~\ref{model}.
From the wavefunction~\eq{eq:flatbandwf} it is possible to explicitly
calculate the spin polarization. For the $x$-spin polarization, the
unwieldy full expression is greatly simplified in the limit
$\kappa_{-} \ll k_{x,-},\, | p_{+} |$, which is realized for $|k_y|$ close to
$k_{F,-}$, where we find
\beq
\rho^x(k_y)  \cong -\frac{k_y}{k_{F,-}} , \label{eq:rhoxflatband}
\eeq
in excellent agreement with the numerics.
Note that this is the
$x$-spin polarization expected for purely negative-helicity states;
indeed, negative-helicity states contribute almost all the weight
of the flat-band wavefunctions for $\kappa_{-}\ll | p_{+} | $, as the
positive-helicity components are sharply localized at the edge.
Significant deviations from~\eq{eq:rhoxflatband} therefore occur when
the localization length for the negative-helicity sector is
comparable or larger than that for positive helicity,
i.e., for $\kappa_{-} \gtrsim  | p_{+} | $, which occurs close to the
projected positive-helicity gap nodes at $k_y=\pm k_{F,+}$.  This is
not surprising, as the edge-state wavefunction must evolve to match
the bulk positive-helicity wavefunctions at the node. Thus, within
the momentum range $|k_y|/k_{F,+}-1
\lesssim ([q-q_c]/\xi_0 k_F)^{2}$ (obscured by the green
lines in figure~\ref{2DC4vdp}), the $x$-spin polarization reverses and  at
$|k_{y}| = k_{F,+} + 0^{+}$ is equal to $\sgn(k_y)$.
This illustrates an important principle: the spin polarization of
the nondegenerate flat bands varies so that it matches the spin
polarization of the bulk positive-helicity states at the
bounding nodes. Since the positive-helicity gap vanishes here, these
are in fact identical to the positive-helicity states in the normal phase.

\section{Surface states of three-dimensional NCSs}~\label{3DNCS}

We now turn to the  case of three-dimensional NCSs.
For simplicity we ignore the spin-orbit splitting of the
Fermi surfaces, i.e., we set $\widetilde{\lambda}=0$. The effects of
the inversion-symmetry breaking is therefore restricted to the mixed
parity of the superconducting gap. We do not expect this to
qualitatively alter our results~\cite{schnyder2012}.
Assuming the $s$-wave
form-factor $f({\bf k})=1$, nodal rings appear only in the negative-helicity
gap for  $0<q<q_c=0.5$; for $q>q_c$ the system is fully gapped and
topologically trivial. Nondegenerate zero-energy flat-band surface
states appear in the surface Brillouin zone for ${\bf k}_\parallel$
lying within the projections of the nodal lines, such that the
negative-helicity gap has opposite sign at ${\bf k}^{}_{-}$ and
${\bf k}^\prime_{-}$, i.e., on opposite sides of the
Fermi surface~\cite{schnyder2011,schnyder2012,brydon2011}.

\begin{figure}[t]
\includegraphics[width=\columnwidth]{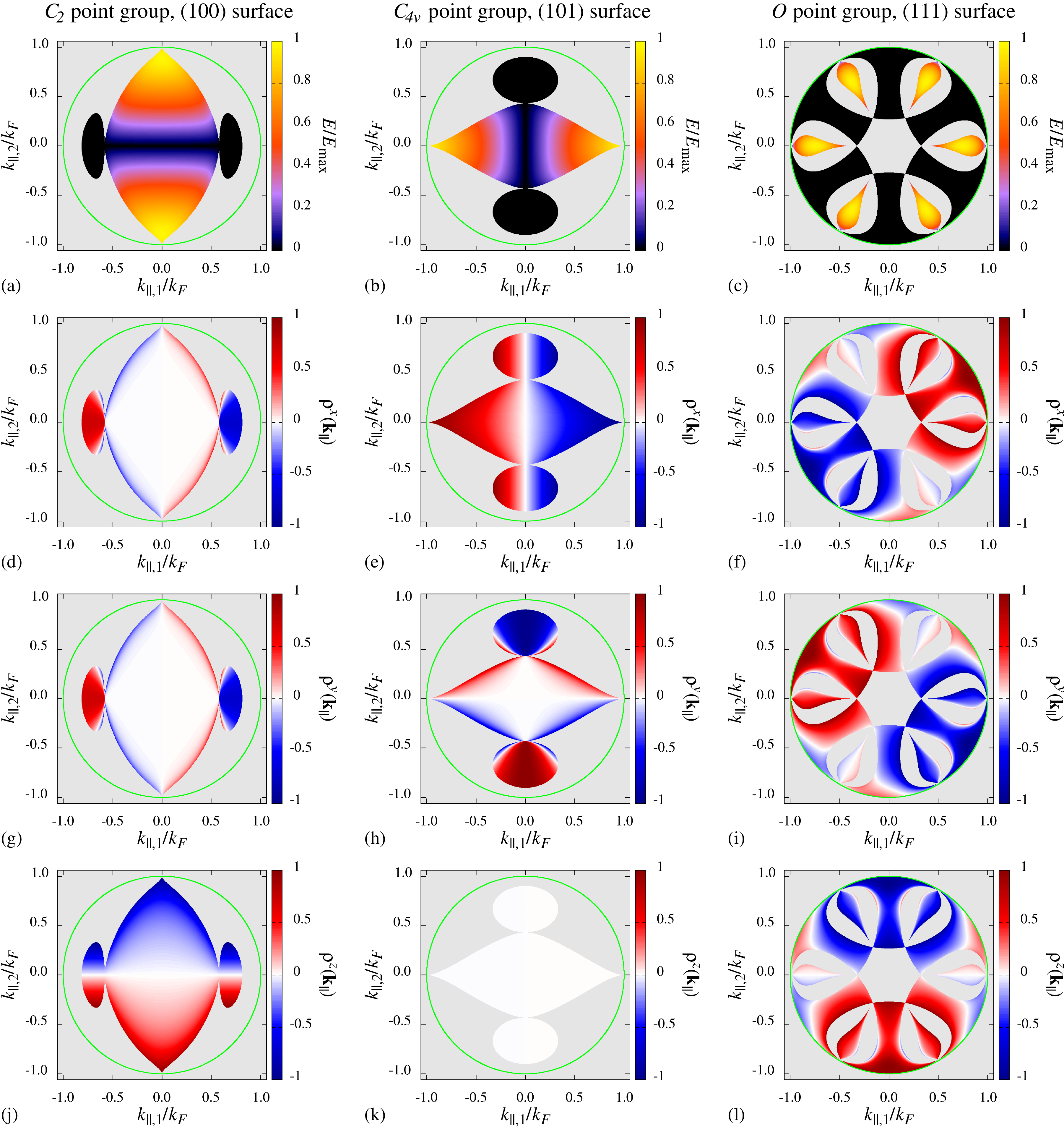}
\caption{\label{3D} Edge-state dispersion and spin polarization of 3D
NCS systems. 
Each column shows results for a different point group
symmetry: from the left we have the (100) surface of a $C_{2}$ NCS,
the (101) surface of a $C_{4v}$ NCS, and the (111) surface of an $O$
NCS. The first row gives the dispersion of the nonnegative-energy edge
states, while the second, third, and fourth rows shows the spin
polarization along 
the $x$, $y$, and $z$ axis of the crystal, respectively. In all  panels
we take $q=0.25$ and assume negligible spin-orbit splitting,
$\widetilde{\lambda}=0$. Grey indicates the absence of any surface
state, black regions in panels (a), (b), (c) denote flat
zero-energy bands, and the green circle is the projection of the Fermi surface.
}
\end{figure}

In figure~\ref{3D} we plot typical dispersions of surface states
for the three point groups
considered here: monoclinic $C_{2}$ in the left column, tetragonal
$C_{4v}$ in the middle column, and cubic $O$ in the right
  column.
In each case the zero-energy flat-band states coexist with dispersing edge
states. Similar to the two-dimensional systems studied above, both the
dispersing and  the flat-band states are generally spin polarized, as
shown in the lower three rows of figure~\ref{3D}. Note that the spin
polarization is given with respect to the crystal axes.

Although the variation of the spin polarization across the flat band
can be rather complicated, we know from the discussion of the
two-dimensional systems that the polarization of the flat-band edge
states must match that of the normal bulk states at the bounding
nodes. Since the gap nodes appear only in the negative-helicity
sector, the spin polarization close to the edge of the flat bands is
$-{\bf l}_{\bf k}/|{\bf l}_{\bf k}|$ where ${\bf k}$ lies on the gap
node. This is nicely illustrated by the case of the $C_{4v}$ NCS,
where the spin polarization of the flat-band states rotates
in a clockwise direction in the  $xy$ plane as one moves around
their edge in the same sense, consistent with the negative helicity of
the normal states at the gap node.

\section{Experimental tests of the spin texture}

The nontrivial spin texture of NCS surface states strongly influences
their surface physics. For example, the opposite sign of the spin
polarization of the surface states on opposite sides of the surface
Brillouin zone
forbids spin-independent scattering between
them~\cite{schnyderPRL2013}. This characteristic property, which
results in a partial protection of the surface states against
localization from nonmagnetic impurities \cite{queirozPRB2014}, can be
observed experimentally using quasiparticle-interference patterns
measured by scanning tunneling
microscopy~\cite{hofmannprb2013}. Another possibility is to probe the
surface-state polarization by bringing the NCS into contact with a
ferromagnet.
In the case of nodal NCSs, the coupling of the flat-band states to the exchange
field of the ferromagnet induces a nonzero edge-state dispersion, thereby
converting the flat bands into chirally dispersing surface modes. This results in
 a surface charge current with a distinctive singular dependence on
the exchange-field strength \cite{brydonNJP2013,schnyderPRL2013}.

Here we examine a complementary approach to measure the spin
polarization of the NCS surface states. Namely, we consider the conductance of a
 tunnel junction between a normal metal and an NCS separated by an insulating
ferromagnetic barrier. In this setup, the magnetization of the
insulating tunnel barrier leads to an energy shift of the NCS surface states,
which in turn changes the tunneling conductance. Thus, as we shall demonstrate
below,  the spin polarization of the surface states
leads to a strong dependence of the zero-bias conductance on the
orientation of the magnetization of the barrier.

We note that the interface physics of NCS-ferromagnet heterostructures
probe only the \emph{local} spin density of the states near the
interface, which cannot be easily related to the surface-state spin
polarization. This could in
principle be evidenced by applying an exchange field to the bulk
NCS~\cite{yadaPRB2011,matsuuraNJP2013,wongPRB2013}, which however
also produces a pronounced reconstruction of the pairing
state~\cite{agterbergPRB2007,loderJPCM2013}. The spatial separation of
the exchange field and the bulk NCS in heterostructure devices avoids
this problem.


We consider a junction between an NCS and a normal metal with a
ferromagnetic insulator as the tunnel barrier. To calculate the
tunneling conductance we use a generalization of the
Blonder-Tinkham-Klapwijk formula~\cite{tanakaNagaosaPRB09,BTK,yokoyamaPRB2005},
\beq
\sigma_{S}(E) = \sum_{{\bf k}_\parallel}\left\{1 +
\frac{1}{2}\sum_{\sigma,\sigma'}\left(|a^{\sigma,\sigma'}_{{\bf
    k}_\parallel}|^2 - |b^{\sigma,\sigma'}_{{\bf k}_\parallel}|^2\right)\right\},
\eeq
where $a^{\sigma,\sigma'}_{{\bf k}_\parallel}$ and
$b^{\sigma,\sigma'}_{{\bf k}_\parallel}$ are the Andreev and normal
reflection coefficients, respectively, for spin-$\sigma$ electrons
injected into the NCS at interface momentum ${\bf k}_\parallel$.  Due
to the magnetic barrier and the spin structure of the NCS, the 
reflected holes and electrons can have spin $\sigma^\prime=\sigma$
and $\sigma^\prime = -\sigma$. The scattering
coefficients are determined by solving the BdG
equations for the junction at energy $E$. An appropriate ansatz for
the scattering wavefunction for an injected spin-$\sigma$ electron is
\beqarray
\fl \qquad  \psi_\sigma({\bf k}_\parallel,{\bf r}) = \psi^{\rm N}_{e,\sigma}e^{i{\bf
    k}\cdot{\bf r}} + \sum_{\sigma'}\left\{a^{\sigma,\sigma'}_{{\bf
    k}_\parallel}\psi^{\rm N}_{h,\sigma'} e^{i{\bf k}\cdot{\bf r}} + b^{\sigma,\sigma'}_{{\bf
    k}_\parallel}\psi^{\rm N}_{e,\sigma'} e^{i{\bf k}^{\prime}\cdot{\bf r}}\right\}\Theta(r_\perp)
\nonumber \\
\qquad  + \sum_{\nu}\left\{c^{\sigma,\nu}_{{\bf k}_\parallel}\psi^{\rm
  NCS}_{e,\nu}({\bf k})e^{i{\bf k}\cdot{\bf r}} + d^{\sigma,\nu}_{{\bf k}_\parallel}\psi^{\rm
  NCS}_{h,\nu}({\bf k}^\prime)e^{i{\bf k}^\prime\cdot{\bf r}}\right\}\Theta(-r_\perp) ,
\eeqarray 
with the wavevectors ${\bf k} = ({\bf k}_\parallel,k_\perp)$ and
${\bf k}^\prime = ({\bf k}_\parallel,-k_\perp)$. For simplicity we
assume that the normal metal and the NCS have the
same Fermi surface radius $k_F$ and effective mass $m$ and we employ
the Andreev approximation, where all wavevectors are assumed to have
magnitude equal to $k_F$. Relaxing these common approximations is
not expected to qualitatively alter our conclusions. The electron and
hole spinors in the normal lead are defined as
\beqarray
\psi^{{\rm N}}_{e,\sigma} = \frac{1}{2}\left(\begin{array}{cccc}1 + \sigma, &
  1-\sigma, & 0, & 0\end{array}\right)^{\!\mathrm{T}}, \\
\psi^{{\rm N}}_{h,\sigma} = \frac{1}{2}\left(\begin{array}{cccc}0, & 0, & 1 + \sigma, &
  1-\sigma\end{array}\right)^{\!\mathrm{T}}
\eeqarray
and the electron- and hole-like spinors in the NCS are given by
\beqarray
\psi^{{\rm NCS}}_{e,\nu} =
\frac{1}{\sqrt{2}}\left(\begin{array}{cccc}u^\nu_{{\bf k}}, & \nu \frac{l^{x}_{\bf k} +
    il^{y}_{\bf k}}{|{\bf l}_{\bf k}| +\nu l^{z}_{\bf k}}u^\nu_{{\bf k}}, & -\nu\frac{l^{x}_{\bf k}
    + il^{y}_{\bf k}}{|{\bf l}_{\bf k}| + \nu l^{z}_{\bf k}}\,
  s^\nu_{{\bf k}}v^\nu_{{\bf k}}, &
  s^\nu_{{\bf k}}v^\nu_{{\bf k}}\end{array}\right)^{\!\mathrm{T}} , \\
\psi^{\rm NCS}_{h,\nu} = \frac{1}{\sqrt{2}}\left(\begin{array}{cccc}v^\nu_{{\bf k}}, & \nu \frac{l^{x}_{\bf k} +
    il^{y}_{\bf k}}{|{\bf l}_{\bf k}| +\nu l^{z}_{\bf k}}v^\nu_{{\bf k}}, & -\nu\frac{l^{x}_{\bf k}
    + il^{y}_{\bf k}}{|{\bf l}_{\bf k}| + \nu l^{z}_{\bf k}}\,
  s^\nu_{{\bf k}}u^\nu_{{\bf k}}, &
  s^\nu_{{\bf k}}u^{\nu}_{{\bf k}}\end{array}\right)^{\!\mathrm{T}} ,
\eeqarray
with $s^\nu_{{\bf k}} = \sgn(\Delta^\nu_{{\bf k}})$ and
\beq
u^\nu_{{\bf k}} = \sqrt{\frac{E + \Omega^\nu_{{\bf k}}}{2E}} , \qquad
v^\nu_{{\bf k}} = \sqrt{\frac{E - \Omega^\nu_{{\bf k}}}{2E}} ,
\eeq
where $\Omega^\nu_{{\bf k}} = \sqrt{E^2 - |\Delta^\nu_{\bf k}|^2}$.

We model the insulating barrier as a $\delta$-function at $r_\perp =
0$, with charge and magnetic potentials $U_c>0$ and
$U_{s}>0$, respectively~\cite{kashiwayaPRB1999}. The wavefunction is
continuous across the barrier,
\beq
\Psi_\sigma({\bf k}_{\parallel},{\bf
  r})|_{r_\perp = 0^+} = \Psi_\sigma({\bf k}_{\parallel},{\bf
  r})|_{r_\perp = 0^-} ,
\eeq
but its derivative is discontinuous,
\beq
\fl \partial_{r_\perp}\!\Psi_\sigma({\bf k}_{\parallel},{\bf
  r})|_{r_\perp = 0^+} -
\partial_{r_\perp}\!\Psi_\sigma({\bf k}_{\parallel},{\bf r})|_{r_\perp = 0^{-}}
 = 2(Z_c +
  Z_{s}\,\hat{\bf M}\cdot[\tau^z\check{\bf S}])\,
  \Psi_\sigma({\bf k}_{\parallel},{\bf r})|_{r_\perp = 0},
\eeq
where $Z_c=mU_c/\hbar^2$, $Z_s = mU_s/\hbar^2$, and $\hat{\bf M}$ is
the unit vector in the direction of the magnetization. We require
$Z_c>Z_s>0$ to describe a ferromagnetic insulator.

\begin{figure}[t]
\begin{center}
\includegraphics[width=0.6\columnwidth]{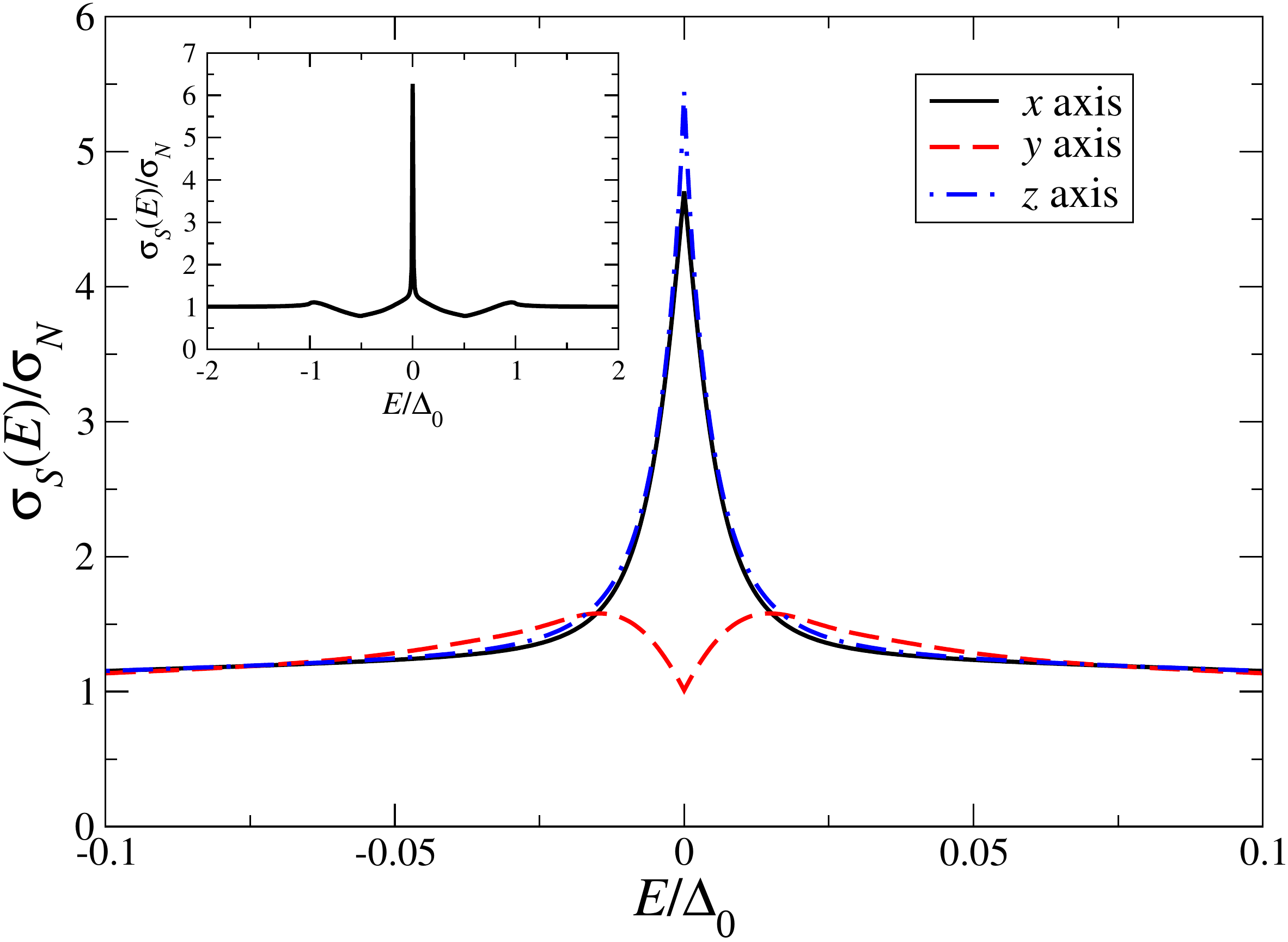}
\end{center}
\caption{\label{conFI} Normalized conductance spectra close to zero
bias at the $(101)$ surface
of a three-dimensional NCS with $C_{4v}$ point group. The curves
show the dependence of the tunneling conductance on the orientation of
the magnetization 
of the insulating tunnel barrier, which is modeled by the
parameters $Z_c=4$ and $Z_s=1$. We ignore the
effect of spin-orbit splitting in the NCS and set $q=0.25$.
Inset: normalized conductance spectrum over a larger range of energies for
a nonmagnetic barrier with $Z_c=4$ and $Z_s=0$; turning on the
magnetic potential does not significantly modify the conductance
spectra outside the energy range in the immediate vicinity of the
zero-bias conduction peak.}
\end{figure}

In figure~\ref{conFI} we plot the conductance normalized by the normal-state
value $\sigma_N$ for tunneling through the (101)
surface of a $C_{4v}$ NCS. As shown in the inset, the conductance
spectrum for a nonmagnetic tunnel barrier is dominated by a sharp peak
at zero bias, which arises from resonant tunneling through the
zero-energy flat-band states, which are now resonances in the
NCS due to the nonzero barrier
transparency~\cite{brydon2011,tanakaPRL2010}. Upon
switching on the barrier magnetization, the peak remains intact for
a magnetization in the $xz$ plane
but disappears completely for a magnetization along the $y$ axis.
The conductance spectrum at larger bias
is essentially unaffected by the barrier magnetization.
This behavior results from the coupling of the
barrier magnetization to the surface spin density of the
nondegenerate flat-band resonances. A naive perturbative argument implies that
the energy of the resonance should be shifted by an amount proportional to
the surface spin polarization. Shifting the resonance away from zero
energy results in a reduction of the conductance peak at zero bias,
which is indeed observed in figure~\ref{conFI}.
We remark that the strong dependence of the tunneling conductance on the
orientation of the barrier magnetization, which is  shown in figure~\ref{conFI},
is qualitatively different from
the behavior of the tunneling conductance
in an equivalent junction involving a singlet $d$-wave superconductor.
In the latter case,  the ferromagnetic tunnel barrier splits the spin-degenerate surface states
of the superconductor for arbitrary orientations of the barrier 
magnetization, which in turn leads to a suppression of the zero-bias
conductance independent of this
orientation~\cite{kashiwayaPRB1999}. 

Although the surface spin density is not equivalent to the total spin
polarization, we nevertheless find the latter to be a good guide to the fate of
the zero-bias peak. Examining the spin polarization for the states at the
(101)  surface of the $C_{4v}$ NCS in figures~\ref{3D}(f)--(h), we
see that the absence of the
zero-bias peak for a $y$-polarized barrier but its presence for a
$z$-polarized barrier is consistent with the strong spin polarization of
the surface states along the $y$ axis and the vanishing spin polarization
along the $z$ axis. Although the height of the zero-bias peak
  for an $x$-polarized barrier is approximately 15\% smaller than for
  the $z$-polarized barrier, the survival of the
zero-bias peak in spite of the strong $x$-spin polarization of the
surface states is somewhat surprising. A possible explanation is that
the states with the strongest $x$-spin polarization have surface
momenta close to the projected nodal lines, and thus have diverging
localization lengths. This strongly suppresses their weight at the
tunnel barrier, and hence their energy shift due to the  coupling to
the barrier moment is also reduced. 

\begin{figure}[t]
\includegraphics[width=\columnwidth]{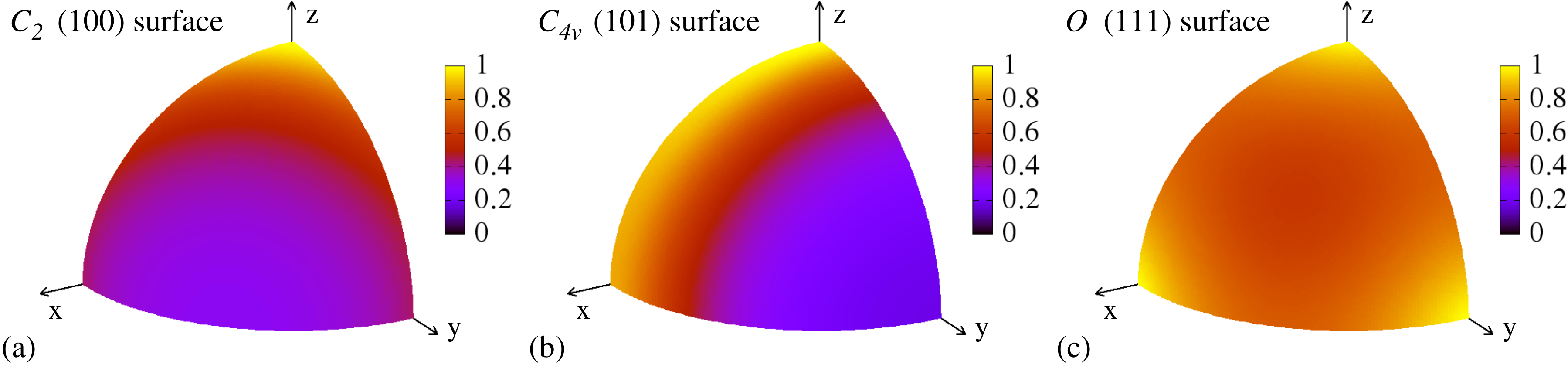}
\caption{\label{zbcpmap} Variation of the zero-bias conductance
as a function of  the orientation of the barrier
magnetization for (a) a $C_2$ NCS at the
(100) surface, (b) a $C_{4v}$ NCS at the (101) surface, and (c) a
$O$ NCS at the (111) surface. The position on the surface of the
one-eighth sphere represents
the orientation of the magnetization, while the color
at each point gives the ratio of the zero-bias conductance to the
maximum  value. In all panels we set $q=0.25$,
$Z_c=4$, and $Z_s=1$.}
\end{figure}

The dependence of the zero-bias conductance on the orientation of the barrier
magnetization is shown in figure~\ref{zbcpmap} for the three NCS point groups
considered in section~\ref{3DNCS}. In all cases we find not only a strong
variation of the conductance as a function of the orientation but also
observe that this dependence is distinctly different for each point
group. This can be exploited to test the existence of spin-polarized
flat bands and also to identify the pairing symmetry of the NCS.

\section{Summary and outlook}

In this work we have presented a systematic study of the spin polarization
of NCS surface states using quasiclassical scattering theory. Examining both
fully gapped and nodal pairing states, we have shown how the spin polarization
generally depends  on the interplay of  spin-orbit coupling and 
singlet-triplet pairing ratio in the superconductor. The variation of the
surface-state spin polarization strongly reflects the relative weight of
negative- and positive-helicity wavefunction components and is to some degree
controlled by the spin polarization of the bulk states at the point where the
surface states connect to the bulk continuum. This is particularly
pronounced in
nodal NCSs, where the spin polarization of the surface states evolves to match
that of the normal states at the gap nodes. We have also shown that the spin
polarization of the surface states can be directly probed in a tunnel
junction consisting of
a normal metal and an NCS separated by an insulating
ferromagnetic barrier. Specifically, the dependence of the zero-bias conductance
on the orientation of the barrier magnetization is a signature of
spin-polarized flat-band surface states.

Our results provide a deeper understanding of the surface
physics of NCS, which reflects the topological properties of these
materials. Although the spin polarization of the surface states is not
directly related to their topology, it can nevertheless be exploited
in experiments to detect the topological surface
states and to probe their degeneracy. We believe that our findings
will prove relevant for designing experiments to test the topological
character of NCS and other unconventional superconductors.
While we have focused in this work on NCSs with one spin-split Fermi surface,
our analysis can be generalized in a straightforward manner to other topological superconductors~\cite{schnyderPRB08},
e.g., centrosymmetric systems with triplet pairing, locally noncentrosymmetric superconductors~\cite{biswasPRB2013,fischerPRB14}, Weyl superconductors~\cite{mengPRB2012}, and superconductors with multiple spin-split Fermi sheets.

\ack
The authors thank M.\ Sigrist, P.\ Wahl, and G.\ Annunziata for useful
discussions. 
C.\,T. acknowledges financial support by the Deutsche
Forschungsgemeinschaft through Research Training Group GRK 1621.

\section*{References}

\bibliography{refsNJP}

\end{document}